\documentclass[twocolumn,showpacs,preprintnumbers,amsmath,amssymb,nofootinbib,APS]{revtex4}

\usepackage{dcolumn}
\usepackage{bm}

\begin{document}

\preprint{}%gr-qc/xxxxxxx

\title{ {\bf Post-Newtonian constraints on $f(R)$ cosmologies in metric formalism  }}
\author{Gonzalo J. Olmo}\email{gonzalo.olmo@uv.es}
\affiliation{ {\footnotesize Departamento de F\'{\i}sica Te\'orica and
    IFIC, Centro Mixto Universidad de
    Valencia-CSIC} \\
    {\footnotesize Facultad de F\'{\i}sica, Universidad de Valencia, Burjassot-46100, Valencia,
    Spain}\\
     {\footnotesize  Physics Department, University of
Wisconsin-Milwaukee,Milwaukee, WI 53201, USA }}

\date{May 26th, 2005}

\begin{abstract}
We compute the complete post-Newtonian limit of the metric form of
$f(R)$ gravities using a scalar-tensor representation. By
comparing the predictions of these theories with laboratory and
solar system experiments, we find a set of inequalities that any
lagrangian $f(R)$ must satisfy. The constraints imposed by those
inequalities allow us to find explicit bounds to the possible
nonlinear terms of the lagrangian. We conclude that the lagrangian
$f(R)$ must be almost linear in $R$ and that corrections that grow
at low curvatures are incompatible with observations. This result
shows that modifications of gravity at very low cosmic densities
cannot be responsible for the observed cosmic speed-up.

\end{abstract}


\maketitle

\section{Introduction}

It is now generally accepted that the universe is undergoing a
period of accelerated expansion \cite{Tonry03,Knop03}, which
cannot be justified by the description provided by the equations
of motion of General Relativity (GR) and a universe filled with
standard sources of matter and energy. It has been suggested that
this effect could have its origin in, among other possibilities,
corrections to the equations of motion of GR generated by
nonlinear contributions of the scalar curvature in the gravity
lagrangian $f(R)$ \cite{CAPO,CDTT,VOL} (see also
\cite{Carroll:2004de}). Reasons for considering nonlinear
curvature terms in the gravity lagrangian can be found in quantum
effects in curved space \cite{PARKER} or in certain low-energy
limits of string/M-theories \cite{N-O}. The nonlinearity of the
lagrangian can also be related to the existence of scalar degrees
of freedom in the gravitational interaction \cite{TT-W}. In any
case, the fact that certain $f(R)$ lagrangians naturally lead to
early-time inflationary behaviors is the main motivation to study
possible new gravitational effects in
the late-time cosmic expansion. \\

Once a nonlinear lagrangian $f(R)$ has been proposed, the
equations of motion for the metric can be derived in two
inequivalent ways. On the one hand, one can follow the standard
metric formalism, in which variation of the action with respect to
the metric leads to a system of fourth-order equations. On the
other hand, one may assume that metric and connection are
independent fields and then take variations of the action with
respect to the metric and with respect to the connection. In this
case, the resulting equations of motion for the metric are
second-order. Only when the function $f(R)$ is linear in $R$, GR
and GR plus cosmological constant, metric and Palatini formalisms
lead to the same equations of motion. In this work we will be
exclusively concerned with the metric formalism. The Palatini
formalism will we considered elsewhere \cite{PII}.\\

Though much work has been carried out in the last few years with
regard to $f(R)$ gravities in the cosmological regime, very little
is known about the form that the gravity lagrangian should have in
order to be compatible with the most recent cosmological
observations \cite{matching-data}. The main reason for this seems
to be the fact that the precision of the supernovae luminosity
distance data and other currently available tests supporting the
late-time cosmic speed-up is not enough to discriminate with
confidence between one model or another. It would be thus
desirable to have a new arena where to test these theories with
higher precision. In our opinion, the solar system represents a
scenario more suitable than the cosmological one to study the
possible constraints on the lagrangian $f(R)$. In fact, if in
addition to modified gravitational dynamics, sources of dark
energy were acting in the cosmic expansion, it would be very
difficult to distinguish their effect from a purely gravitational
one. In the solar system, however, it is ordinary matter which
dominates the gravitational dynamics, being the contribution of
dark sources negligible. Therefore, we should see the solar system
as a suitable laboratory to impose the first useful constraints on $f(R)$ cosmologies. \\

In order to confront the predictions of a given gravity theory
with experiment in the solar system, it is necessary to compute
its weak-field, slow-motion (or post-Newtonian) limit. This limit
has been computed for many metric theories of gravity and put in a
standardized form \cite{WILL}, which depends on a set of
parameters that change from theory to theory (Parametrized
Post-Newtonian [PPN] formalism). However, for $f(R)$ gravities
this limit has not yet been computed in detail. The Newtonian
limit of these theories was recently studied in \cite{DICK}. On
the other hand, since the metric form of $f(R)$ gravities can be
represented as the case $\omega =0$ of Brans-Dicke-like
scalar-tensor theories, it is tempting to use the post-Newtonian
limit of those theories given in the literature \cite{WILL,WAG}
(see also \cite{NI}) to check the viability of particular models.
This was proposed in \cite{CHI}, where it was concluded that the
Carroll et al. model \cite{CDTT} (in metric formalism) was ruled
out according to the observational constraints on the parameter
$\gamma$ corresponding to that model. That result was based on the
fact that the scalar field had a small effective mass, which was
computed in terms of the second derivative of the potential.
However, that prescription is usually derived under the assumption
that the potential and its first derivative vanish (see for
instance \cite{WAG,S-W}), conditions that, in general, cannot be
imposed on $f(R)$ theories (see section \ref{sec:boundary}).\\

In this work we will use the scalar-tensor representation
introduced in \cite{CHI} to compute the post-Newtonian limit of
$f(R)$ gravities taking into account all the terms associated to
the potential of the scalar field. In other words, we will not
make any assumption or simplification about the function $f(R)$
that defines the lagrangian. We will actually compute the
post-Newtonian limit corresponding to Brans-Dicke-like
scalar-tensor theories with arbitrary potential and a generic
constant value of $\omega $ and will then particularize to the
case $\omega =0$, which corresponds to the metric form of $f(R)$
gravities. In this manner we generalize the results of the
literature so as to include all the terms that are relevant for
our discussion. The Palatini form of $f(R)$ gravities, which can
be represented as the case $\omega =-3/2$ of Brans-Dicke-like
theories, represents an exception of the general case $\omega =$
constant and will be studied elsewhere \cite{PII}.\\

The resulting post-Newtonian metric will allow us to confront the
predictions of these theories with the observational data. In this
way, we will find a series of constraints for the lagrangian,
which is a priori completely unknown. Those constraints turn out
to be so strong that the lagrangians compatible with observations
are bounded by a well defined function that prohibits the growing
of the nonlinear terms at low curvatures. This result will be
enough to invalidate the arguments supporting the cosmic speed-up
as due to new gravitational effects at low curvatures.\\

The paper is organized as follows. We first derive the equations
of motion in the original $f(R)$ form and show how to obtain the
scalar-tensor representation out of them. Then we comment on the
choice of coordinates and boundary conditions. The post-Newtonian
limit is computed in the Appendix and commented in section
\ref{sec:PN-metric}, where we obtain constraints on the lagrangian
from the experimental data. Section \ref{sec:examples} is devoted
to the discussion of particular models. In section \ref{sec:cons}
we find the form of the lagrangian that satisfies the constraints.
We conclude the paper with a brief summary and conclusions.

\section{ Equations of motion }\label{sec:eom}

The action that defines $f(R)$ gravities in the metric formalism
is the following
\begin{equation}\label{eq:def-f(R)}
S[f;g,\psi_m]=\frac{1}{2\kappa ^2}\int d^4
x\sqrt{-g}f(R)+S_m[g_{\mu \nu},\psi_m]
\end{equation}
where $S_m[g,\psi_m]$ represents the matter action, which depends
on the metric $g_{\mu \nu }$ and the matter fields $\psi_m$. For
notational purposes, we remark that $T_{\mu \nu
}=\frac{-2}{\sqrt{-g}}\frac{\delta S_m}{\delta g^{\mu \nu }}$, and
that the scalar curvature $R$ is defined as the contraction
$R=g^{\mu \nu }R_{\mu \nu }$, where $R_{\mu \nu }$ is the Ricci
tensor
\begin{equation}\label{eq:def-Ricci}
R_{\mu\nu}=-\partial_{\mu}
\Gamma^{\lambda}_{\lambda\nu}+\partial_{\lambda}
\Gamma^{\lambda}_{\mu\nu}+\Gamma^{\lambda}_{\mu\nu}\Gamma^{\rho}_{\rho\lambda}-\Gamma^{\lambda}_{\nu\rho}\Gamma^{\rho}_{\mu\lambda}
\end{equation}
and $\Gamma^\alpha_{\beta \gamma }$ is the Levi-Civita connection
\begin{equation}\label{eq:def-Gamma}
\Gamma^\alpha_{\beta \gamma }=\frac{g^{\alpha \lambda
}}{2}\left(\partial_\beta g_{\lambda \gamma }+\partial_\gamma
g_{\lambda \beta }-\partial_\lambda g_{\beta \gamma }\right)
\end{equation}
From eq.(\ref{eq:def-f(R)}) we obtain the following equations of
motion:
\begin{equation}\label{eq:f-var}
f'(R)R_{\mu\nu}-\frac{1}{2}f(R)g_{\mu\nu}-
\nabla_{\mu}\nabla_{\nu}f'(R)+g_{\mu\nu}\Box f'(R)=\kappa ^2T_{\mu
\nu }
\end{equation}
where $f'(R)\equiv df/dR$. According to eq.(\ref{eq:f-var}), we
see that, in general, the metric satisfies a system of
fourth-order partial differential equations. The higher order
derivatives come from the terms $\nabla_{\mu}\nabla_{\nu}f'$ and
$\Box f'$. Only when $f(R)$ is a linear function of the scalar
curvature, $f(R)=a+bR$, the equations of motion are second-order.
The trace of eq.(\ref{eq:f-var}) is given by
\begin{equation}\label{eq:f-var-trace}
3\Box f'+f'R-2f=\kappa ^2T
\end{equation}
This equation can be interpreted as the equation of motion of a
self-interacting scalar field, where the self-interaction terms
are represented by $f'R-2f$. This can be seen by algebraically
inverting the function $f'(R)$ and writing $R$ as $R=R(f')$. In
this way, defining
\begin{eqnarray}\label{eq:phi=f'}
\phi&\equiv& f'\\
V(\phi)&\equiv& R(\phi)f'-f(\phi) \label{eq:V=rf'-f}
\end{eqnarray}
we can write eqs.(\ref{eq:f-var}) and (\ref{eq:f-var-trace}) as
\begin{eqnarray}\label{eq:f-var-st}
G_{\mu \nu }&=&\frac{\kappa ^2}{\phi}{T}_{\mu \nu
}-\frac{V(\phi)}{2\phi}{g}_{\mu \nu
}+\frac{1}{\phi}\left({\nabla}_\mu {\nabla}_\nu \phi-{g}_{\mu \nu
}{\Box}\phi\right)\\
3\Box \phi &+& 2V(\phi)-\phi\frac{dV}{d\phi}=\kappa ^2 T
\label{eq:Box-f}
\end{eqnarray}
The above equations of motion can also be obtained from the
following action
\begin{eqnarray} \label{eq:ST}
S[{g}_{\mu \nu},\phi,\psi_m]&=&\frac{1}{2\kappa ^2 }\int d^4
x\sqrt{-{g}}\left[\phi {R}({g})-\right.
\\&-&\left.\frac{w(\phi)}{\phi}(\partial_\mu \phi\partial^\mu\phi)-V(\phi)
\right]+S_m[{g}_{\mu \nu},\psi_m]\nonumber
\end{eqnarray}
which represents a Brans-Dicke-like scalar-tensor theory, in the
particular case $\omega =0$. For more details on the scalar-tensor
representation and a different derivation of this result see \cite{P0}.\\

\section{Coordinates and boundary conditions \label{sec:boundary}}

In order to obtain the metric in the solar system we will follow
the basic guidelines outlined in chapter 4 of Will's book
\cite{WILL}. First we solve eqs.(\ref{eq:f-var-st}) and
(\ref{eq:Box-f}) for the metric and the scalar field in the cosmic
regime, where the high degree of homogeneity and isotropy leads to
a Friedman-Robertson-Walker metric
\begin{equation}\label{eq:FRW}
ds^2=g^B_{\mu \nu }dx^\mu dx^\nu =-dt^2+a(t)^2dx_idx^i
\end{equation}
and to a space-independent value of the scalar field,
$\phi=\phi^B(t) $. At smaller scales there appear local deviations
from the cosmic values of the fields. In the solar system, for
instance, the local deviations will be small, thereby allowing us
to treat them as a perturbation with respect to the background
cosmic boundary values $g^B_{\mu \nu }$ and $\phi ^B(t)$. In our
computations we will use coordinates $(\bar{t},\bar{x}^j)$ in
which the outer regions of the local system are in free fall with
respect to the surrounding cosmological model. Neglecting the
second-order corrections, the local and background coordinates are
simply related by $\bar{t}(t_0,x_0;t,x)\approx (t-t_0)$ and
$\bar{x} ^j(t_0,x_0;t,x)\approx a_0(x-x_0)^j$. From now on we will
omit the bar on the local coordinates and will denote $\phi_0$,
$\dot{\phi }_0$ the asymptotic boundary values of the scalar field
at the cosmic
time $t_0$, i.e., $\phi _0=\phi^B(t_0)$ and $\dot{\phi} _0=\dot{\phi}^B(t_0)$.\\

For approximately static solutions, corresponding to gravitating
masses such as the Sun or Earth, to lowest-order, we can drop the
terms involving time derivatives from the equations of motion. In
our local coordinate system, the metric can be expanded about its
Minkowskian value as $g_{\mu \nu }= \eta_{\mu \nu }+h_{\mu \nu }$.
The solution for the scalar field can be expressed in the form
$\phi =\phi_0+\varphi (t,x)$, where $\phi _0\equiv\phi (t_0)$ is
the asymptotic cosmic value, which is a slowly-varying function of
the cosmic time $t_0$, and $\varphi (t,x)$ represents the local
deviation from $\phi _0$, which vanishes far from the local
system. We want to remark that since $\phi _0$ and $\dot{\phi }_0$
depend on $R_0$ and $\dot{R}_0$, the metric of the local
post-Newtonian system will also depend on the background cosmic
values $R_0$ and $\dot{R}_0$. The dependence on these background
quantities will make the metric change adiabatically in a cosmic
timescale. This adiabatic evolution could make a theory be
compatible with the current experimental tests during some cosmic
era but fail in other periods. We will give below some examples to
illustrate this effect. In particular, we will comment
on the Carroll et al. $1/R$ model \cite{CDTT}.  \\

With regard to the potential defined for the scalar field, see
eq.(\ref{eq:V=rf'-f}), it is easy to see that $dV/d\phi =R$
\cite{P0}. Since, the curvature can be expressed as
$R=R_0+\mathcal{R}(t,x)$, where $\mathcal{R}(t,x)$ denotes the
local deviation from the background cosmic value $R_0$, it is easy
to see that the scalar field will not, in general, satisfy the
extremum condition $dV/d\phi =0$. This is to be contrasted with
the results of the literature regarding the post-Newtonian limit
of Brans-Dicke-like theories, where it is generally assumed that
the field is near an extremum \cite{WAG,S-W}. We thus see that for
$f(R)$ gravities (or $\omega =0$ Brans-Dicke-like theories) it is
necessary to consider all the terms associated to the potential.
This slight complication, on the other hand, has its own
advantages, since at the end of the calculations we may ask Nature
about the admissible forms that the lagrangian $f(R)$ may have.

\section{Post-Newtonian metric}\label{sec:PN-metric}

As we advanced above, we will expand the equations of motion
around the background values of the metric and the scalar field.
In particular, we will take $g_{\mu \nu }\approx \eta_{\mu \nu
}+h_{\mu \nu }$, $g^{\mu \nu }\approx \eta^{\mu \nu }-h^{\mu \nu
}$, $\phi=\phi _0+\varphi(t,x)$ and $V(\phi )\approx V_0+\varphi
V_0'+\varphi^2 V_0''/2+\ldots$ The complete post-Newtonian limit
needs the different components of the metric and the scalar field
 evaluated to the following orders $g_{00}\sim O(2)+O(4)$,
$g_{0j}\sim O(3)$, $g_{ij}\sim O(2)$ and  $\phi \sim O(2)+ O(4)$
(see \cite{WILL}). The details of the calculations and the
complete post-Newtonian limit for the theories defined in
eq.(\ref{eq:ST}) can be found in the Appendix. For convenience, we
will discuss here only the lowest order corrections, $g_{00}\sim
O(2)$, $g_{ij}\sim O(2)$ and  $\phi \sim O(2)$, of the case we are
interested in, say, $\omega =0$. The order of approximation will
be denoted by a superindex. This approximation will be enough to
place tight constraints on the gravity lagrangian. To this order,
the metric satisfies the following equations
\begin{eqnarray}
-\frac{1}{2}\nabla^2\left[h^{(2)}_{00}-\frac{\varphi^{(2)} }{\phi
_0}\right]&=&\frac{\kappa ^2\rho }{2\phi
_0}+\left(\frac{3}{2}\frac{\ddot{\phi }_0}{\phi _0}-\frac{V_0}{2\phi _0}\right)\\
-\frac{1}{2}\nabla^2\left[h^{(2)}_{ij}+\delta
_{ij}\frac{\varphi^{(2)} }{\phi _0}\right]&=&\delta
_{ij}\left[\frac{\kappa ^2\rho}{2\phi _0} -\frac{\ddot{\phi
}_0}{2\phi _0}+\frac{V_0}{2\phi _0}\right]
\end{eqnarray}
where the gauge condition $h^\mu _{k,\mu }-\frac{1}{2}h^\mu _{\mu
,k}=\partial_k\varphi^{(2)} /\phi _0$ has been used. In
eliminating the zeroth-order terms in the field equation for
$\varphi$, corresponding to the cosmological solution for $\phi
_0$, the equation for the scalar field to this order boils down to
\begin{equation}
\left[\nabla^2-m_\varphi^2\right]\varphi^{(2)} (t,x)=-\frac{\kappa
^2\rho }{3}
\end{equation}
where $m^2_\varphi$ is a slowly-varying function of the
cosmological time given by
\begin{equation} \label{eq:m-phi}
m^2_\varphi \equiv \frac{\phi _0 V''_0-V'_0}{3}
\end{equation}
Note that, despite our notation, there is no a priori restriction
on the sign of $m_\varphi ^2$. The  equations of above can be
easily integrated to give
\begin{eqnarray}\label{eq:phi-formal}
\varphi^{(2)} (t,x)&=& \frac{\kappa ^2}{3}\frac{1}{4\pi }\int
d^3x'
\frac{\rho (t,x')}{|x-x'|}F(|x-x'|)\\
h^{(2)}_{00}(t,x)&=& \frac{\kappa ^2}{\phi _0}\frac{1}{4\pi }\int
d^3x' \frac{\rho
(t,x')}{|x-x'|}\left[1+\frac{F(|x-x'|)}{3}\right]-\nonumber \\ &-&
\left(\frac{3}{2}\frac{\ddot{\phi
}_0}{\phi _0}-\frac{V_0}{2\phi _0}\right)\frac{|x-x_c|^2}{3}  \label{eq:h00-formal}\\
h^{(2)}_{ij}(t,x)&=& \left(\frac{\kappa ^2}{\phi _0}\frac{1}{4\pi
}\int d^3x' \frac{\rho
(t,x')}{|x-x'|}\left[1-\frac{F(|x-x'|)}{3}\right]+\right.\nonumber \\
&+& \left.\left[\frac{\ddot{\phi }_0}{2\phi
_0}-\frac{V_0}{2\phi_0}\right]\frac{|x-x_c|^2}{3}\right)\delta
_{ij} \label{eq:hij-formal}
\end{eqnarray}
where $x_c$ is an arbitrary constant vector\footnote{This vector
could be taken, for instance, as representing the center of mass
of the system.} and the function $F(|x-x'|)$ is given by
\begin{eqnarray}\label{eq:Fxx'}
F(|x-x'|)&=&\left\{\begin{array}{cc}
            e^{-m_\varphi |x-x'|} & \makebox{ if } \ m_\varphi
            ^2>0\\ \vspace{0.25cm} \\
            \cos (m_\varphi|x-x'|)  & \makebox{ if } \ m_\varphi ^2<0
            \end{array}\right.
\end{eqnarray}
Note that the term $\dot{\phi }_0$ does not appear in
eqs.(\ref{eq:phi-formal}),(\ref{eq:h00-formal}) and
(\ref{eq:hij-formal}) and, therefore, the fact that $\dot{R}_0$
may not be strictly zero does not affect the Newtonian limit
\cite{DICK}. In the post-Newtonian limit it contributes to
$h^{(4)}_{00}$ (see the Appendix). In any case, since to all
effects $\phi _0$ is almost constant, we can neglect the
contributions due to $\dot{\phi _0}$ and $\ddot{\phi _0}$.\\

Since in the solar system the sun represents the main contribution
to the metric, we can approximate the expressions of above far
from the sources by
\begin{eqnarray}\label{eq:h00}
h^{(2)}_{00}&\approx& 2G\frac{M_{\odot}}{r}+\frac{V_0}{6\phi _0}r^2\\
h^{(2)}_{ij}&\approx& \delta _{ij}\left[2\gamma
G\frac{M_{\odot}}{r}-\frac{V_0}{6\phi _0}r^2\right]\label{eq:hij}
\end{eqnarray}
where $M_{\odot}=\int d^3x'\rho_{sun} (t,x')$ is the Newtonian
mass of the sun and the $\ddot{\phi }_0$ contribution has been
neglected for simplicity. We have defined the effective Newton's
constant $G$ as
\begin{equation}\label{eq:eff-G}
G=\frac{\kappa ^2}{8\pi \phi _0}\left[1+\frac{F(r)}{3}\right]
\end{equation}
and the effective PPN parameter $\gamma $ as
\begin{equation}\label{eq:eff-gamma}
\gamma =\frac{3-F(r)}{3+F(r)}
\end{equation}

We shall show now that the oscillatory solutions, $m_\varphi ^2<0
\ \to \ F(r)=\cos(m_\varphi r)$, are always unphysical. For this
case, the inverse-square law gets modified as follows
\begin{equation}\label{eq:m2<0}
\frac{M_\odot}{r^2} \to \left(1+\frac{\cos(m_\varphi r)+(m_\varphi
r) \sin (m_\varphi r)}{2}\right)\frac{M_\odot}{r^2}
\end{equation}
For very light fields, which represent long-range interactions,
the argument of the sinus and the cosinus is very small in solar
system scales ($m_\varphi r\ll 1$). We can thus approximate
$\cos(m_\varphi r)\approx 1$ and $\sin(m_\varphi r)\approx 0$ and
recover the usual Newtonian limit up to an irrelevant redefinition
of Newton's constant. However, these approximations also lead to
$\gamma \approx 1/2$, which is observationally unacceptable since
$\gamma_{obs}\approx 1$ \cite{WIL-liv}. If the scalar interaction
were short- or mid-range, the Newtonian limit would get
dramatically modified. In fact, the leading order term is then
oscillating, $\sin (m_\varphi r)M_\odot/r$, and is clearly
incompatible with observations. We are thus led to consider only
the damped solutions $F(r)=e^{-m_\varphi r}$.\\

The Yukawa-type correction in the Newtonian potential has not been
observed over distances that range from meters to planetary
scales. In addition, since the post-Newtonian parameter $\gamma $
is observationally very close to unity, we see that the effective
mass in eqs.(\ref{eq:eff-gamma}) and (\ref{eq:eff-G}) must satisfy
the constraint $m_\varphi^2 L^2\gg 1$, where $L$ represents a
typical experimental length scale. Note that when $\omega $ is not
fixed (see the Appendix), there is also the possibility of having
a very light (long-range) field that yields almost
space-independent values of $G$ and $\gamma $. In that case, the
theory behaves as a Brans-Dicke theory with $\gamma $ given by
\begin{equation}
\gamma =\frac{1+\omega }{2+\omega }
\end{equation}
and it takes $\omega>40000$ to satisfy the observational constraints \cite{BIT}.\\

The cosmological constant term $(V_0/6\phi _0)r^2$ appearing in
eqs.(\ref{eq:h00}) and (\ref{eq:hij}) also imposes constraints on
the particular model, since this contribution must be very small
in order not to modify the gravitational dynamics of local systems
ranging from the solar system to clusters of galaxies. In the
terminology of $f(R)$ gravities, the constraint from the
cosmological constant term is
\begin{equation}\label{eq:c2}
\left|\frac{f_0-R_0f'_0}{f_0'}\right|L_L^2\ll 1
\end{equation}
where $L_L$ may represent a (relatively large) length scale the
same order or greater than the solar system. The constraint
$m_\varphi^2 L_S^2\gg 1$ associated to the effective mass can be
reexpressed as
\begin{equation}\label{eq:c1}
\left(\frac{f_0'-R_0f_0''}{f_0''}\right)L^2_S\gg 1
\end{equation}
where $L_S$ represents a (relatively short) length scale that can
range from meters to planetary scales, depending on the particular
test used to verify the theory. It is worth noting that a generic
lagrangian of the form
\begin{equation}\label{eq:ansatz}
f(R)=R+\lambda h(R)
\end{equation}
with $\lambda $ a suitable small parameter, satisfies the two
constraints of above if $h(R)$, $h'(R)$ or $h''(R)$ are finite or
vanish as the universe expands.  General Relativity, which can be
seen as the limit $\lambda \to 0$,  saturates those constraints.
We will consider in the next section some examples of theories
with the form proposed in eq.(\ref{eq:ansatz}). In section
\ref{sec:cons} we will analyze in detail the implications
of the constraint of eq.(\ref{eq:c1}).\\

Before concluding this section, we shall briefly discuss some
simplifications that we may carry out from the above
considerations in the complete post-Newtonian metric given in the
Appendix. First of all, it is worth noting that with a tiny $V_0$
we can eliminate part of the cosmological constant terms. This
fact together with our definition for $G$ leads to the PPN
parameter $\beta=1$, which coincides with the one corresponding to
GR. On the other hand, a massive field would allow us to neglect
the exponential terms and the $\varphi ^{(2)}$ contributions.
Further simplifications could be achieved from the observational
evidence supporting the constancy of Newton's constant. Assuming a
massive field, it follows that $\dot{G}/G\approx -\dot{\phi
}_0/\phi _0$. This relation provides a justification to argue that
$\dot{\phi}_0/\phi _0$ and $\ddot{\phi }_0/\phi _0$ are small, if
nonzero. With these simplifications we recover the post-Newtonian
limit of GR, where $\dot{G}/G=0$. We thus see that measurements of
a change in $G$ with time and of Yukawa-type corrections in the
inverse-square law could be due to the presence of nonlinear
elements in the gravity lagrangian.

\section{Examples}\label{sec:examples}

We shall now illustrate with some simple examples how the
parameters that define the post-Newtonian metric are subject to a
slow adiabatic evolution due to the cosmic expansion. The aim of
this section is to point out the relevance of the cosmic boundary
values of the fields in the description of isolated systems. We
want to make special emphasis on the fact that the gravitational
dynamical properties of a local system at a given time may not be
completely determined by its own internal characteristics, but can
be affected by the state of the universe as a whole at that
moment. Only if the $f(R)$ lagrangian is linear in $R$ or if the
scalar field is non-dynamical (Palatini formalism), the
post-Newtonian metric is completely determined by the properties of the local system.\\

\subsection{Positive powers of R }

Following the structure of the ansatz proposed in
eq.(\ref{eq:ansatz}), we can consider the  family of models
defined by $f(R)=R+R^n/M^{2n-2}$, where $M$ represents a very
large mass scale.  We will only consider the cases $n\ge 2$. These
models are characterized by
\begin{eqnarray}
\phi &\equiv& f' =1+n\left(\frac{R}{M^2}\right)^{n-1} \\
V(\phi)&=&M^2
(n-1)\left(\frac{R}{M^2}\right)^{n}=M^2(n-1)\left(\frac{\phi
-1}{n}\right)^{\frac{n}{n-1}} \label{eq:V-pos}
\end{eqnarray}
With eq.(\ref{eq:V-pos}) at hand, we can compute the effective
mass $m_\varphi ^2$ that characterizes the post-Newtonian metric.
It is given by
\begin{eqnarray}\label{eq:mass-pos}
m_\varphi
^2&=&\frac{R_0}{3(n-1)}\left[\frac{1}{n}\left(\frac{M^2}{R_0}\right)^{n-1}-(n-2)\right]= \nonumber\\
&=& \frac{M^2}{3n}\left(\frac{n}{\phi_0
-1}\right)^{\frac{n-2}{n-1}}\left[1-\frac{(n-2)}{(n-1)}\phi_0
\right]
\end{eqnarray}
where $\phi _0\equiv f'(R_0)$ and $R_0$ represent the cosmological
values of $\phi $ and $R$ at the moment $t_0$. The time-time
component of eq.(\ref{eq:f-var}) can be used to extract some
information about the cosmological evolution of $R$. This will
help us to understand the adiabatic change in the post-Newtonian
metric. The expansion factor satisfies the following equation
\begin{eqnarray}\label{eq:tt}
3\left(\frac{\dot{a}}{a}\right)^2&=&\kappa ^2\rho-
\left(\frac{R}{M^2}\right)^{n-1}\left[3n\left(\frac{\dot{a}}{a}\right)^2-\right.\nonumber
\\&-& \left.\frac{(n-1)}{2}R+
3n(n-1)\frac{\dot{a}}{a}\frac{\dot{R}}{R}\right]
\end{eqnarray}
Inserting $a(t)=a_0e^{\gamma t}$ in eq.(\ref{eq:tt}) and taking
$\rho =0$ for simplicity, it follows that at early-times the
evolution is dominated by the $(R/M^2)^n$ contribution with
$\gamma ^{2(n-1)}=(M^2/12)^{n-1}/(n-2)$. After the early-time
inflation predicted by these expansion factors, as the curvature
decays below the scale defined by $M^2$, the $(R/M^2)^n$ effect is
suppressed and the subsequent evolution is governed by GR, with
$a(t)=a_0t^s$ and $s=1/2$ during the radiation dominated era, and
$s=2/3$ during the matter dominated era. Thus, at all times after
the inflationary period, we have $M^2/R\gg 1$, or equivalently
$(\phi -1)\to 0$. This leads to a very large effective mass for
the scalar field  and a tiny cosmological constant term $V_0/\phi
_0\to 0$. In consequence, this family of models yields an
acceptable weak-field limit. In fact, it seems reasonable to think
that these theories are compatible with GR in all astrophysical
applications, since the curvature is expected to be much smaller
than $M^2$ in all situations but at the very early universe.

\subsection{ Negative powers of R}

\noindent A well-known example of this type is the Carroll et al.
model \cite{CDTT}, defined by $f(R)=R-\mu ^4/R$, where $\mu $
represents a tiny mass scale of order $10^{-33}$ eV. The reason
for the minus sign in front of $\mu ^4$ is intriguing, since this
definition leads to a negative effective mass
\begin{equation}
m_\varphi^2= -\frac{R}{6\mu ^4}(R^2+3\mu ^4)
\end{equation}
which we have shown to be in conflict with the post-Newtonian
limit (see eq.(\ref{eq:m2<0})). An improved formulation of the
theory could be obtained by changing the sign in front of $\mu ^4$
in the definition of $f(R)$. In this way, we can easily extend the
results of the examples of above to the models $f(R)=R+\mu
^{2n+2}/R^n$. A direct consequence of the positive sign in front
of $\mu ^{2n+2}$ is the loss of exponential solutions for $a(t)$
at late-times, since the relation between $\gamma $ and $\mu$
turns into $\gamma ^{2(n+1)}=-(n+2)(\mu ^{2}/12)^{n+1}$. These
models are characterized by
\begin{eqnarray}
\phi &\equiv& f' =1-n\left(\frac{\mu ^2}{R}\right)^{n+1} \label{eq:phi-neg}\\
V(\phi)&=&-\mu^2 (n-1)\left(\frac{\mu
^2}{R}\right)^{n}=\nonumber\\&=&\mu
^2(n+1)\left(\frac{n}{1-\phi}\right)^{\frac{n}{n+1}}
\label{eq:V-neg}
\end{eqnarray}
The effective mass of the scalar field takes the form
\begin{eqnarray}\label{eq:mass-neg}
m_\varphi
^2&=&\frac{R_0}{3(n+1)}\left[\frac{1}{n}\left(\frac{R_0}{\mu
^2}\right)^{n+1}-(n+2)\right]=\nonumber \\ &=&\frac{\mu
^2}{3n}\left(\frac{n}{1-\phi_0
}\right)^{\frac{n+2}{n+1}}\left[\frac{(n+2)}{(n+1)}\phi_0
-1\right]
\end{eqnarray}
We will restrict our discussion to the cases with $n \ge 1$. The
cosmological evolution of these models during the radiation
dominated era requires a complete solution of the model, since a
simple power law expansion is ill-defined. We will just
concentrate on the matter dominated era, $a(t)=a_0t^{2/3}$, and
beyond, $a(t)=\tilde{a_0} t^{s_n}$ with $s_n=(2n+1)(n+1)/(n+2)$.
These solutions imply that the curvature decays with the cosmic
time as $R=6s(2s-1)/t^2$. One can numerically check that the
transition from the matter dominated era, $s=2/3$, to its final
value $s_n$ is smooth (we took $\kappa ^2\rho_{m_0}/\mu ^2=3/7$).
During the matter dominated era, $\mu ^2/R\to 0$ and $\phi \approx
1$, eqs.(\ref{eq:mass-neg}) and (\ref{eq:V-neg}) indicate that
$m_\varphi ^2$ is very large and $V_0/\phi _0$ very small. In
consequence, these models yield a valid post-Newtonian limit.
However, as the universe expands and the curvature approaches the
critical value $(R_c/\mu ^2)^{n+1}=n(n+2)$, in which $m_\varphi
^2=0$, the effective mass is small and the  post-Newtonian limit
tends to that of a Brans-Dicke theory with $\omega =0$. At later
times, $m_\varphi ^2$ becomes negative and the weak-field
approximation is ill-defined, as we discussed above.  We can,
thus, conclude that these theories do not seem a good alternative
to explain the late-time cosmic speed-up, since they have an
unacceptable weak-field limit at the present time.\\

\section{Constrained lagrangian}\label{sec:cons}

A qualitative analysis of the constraint given in eq.(\ref{eq:c1})
can be used to argue that, in general, $f(R)$ gravities with terms
that become dominant at low cosmic curvatures are not viable
theories in solar system scales  and, therefore, cannot represent
an acceptable mechanism for the cosmic expansion. Roughly
speaking, eq.(\ref{eq:c1}) says that the smaller the term $f''_0$,
with $f''_0>0$ to guarantee $m_\varphi ^2>0$, the heavier the
scalar field\footnote{ Note that $\phi \equiv f'$ must be positive
in order to have a well-posed theory.}. In other words, the
smaller $f''_0$, the shorter the interaction range of the field.
In the limit $f''_0\to 0$, corresponding to GR, the scalar
interaction is completely suppressed. Thus, if the nonlinearity of
the gravity lagrangian had become dominant in the last few
billions of years (at low cosmic curvatures), the scalar field
interaction range would have increased accordingly. In
consequence, gravitating systems such as the solar system,
globular clusters, galaxies,\ldots would have experienced (or will
experience) observable changes in their gravitational dynamics.
Since there is no experimental evidence supporting such a change
and all currently available solar system gravitational experiments
are compatible with GR,  it seems unlikely that the nonlinear
corrections may be dominant at the current epoch.\\

Let us now analyze in detail the constraint given in
eq.(\ref{eq:c1}). That equation can be rewritten as follows
\begin{equation}\label{eq:c11}
R_0\left[\frac{f'(R_0)}{R_0f''(R_0)}-1\right]L_S^2\gg 1
\end{equation}
We are interested in the form of the lagrangian at intermediate
and low cosmic curvatures, i.e., from the matter dominated to the
vacuum dominated eras. We shall now demand that the interaction
range of the scalar field remains as short as today or decreases
with time so as to avoid dramatic modifications of the
gravitational dynamics in post-Newtonian systems with the cosmic
expansion. This can be implemented imposing
\begin{equation}\label{eq:difeq-0}
\left[\frac{f'(R)}{Rf''(R)}-1\right]\ge \frac{1}{l^2R}
\end{equation}
as $R\to 0$, where $l^2\ll L_S^2$ represents a bound to the
current interaction range of the scalar field. Thus,
eq.(\ref{eq:difeq-0}) means that the interaction range of the
field must decrease or remain short, $\sim l^2$, with the
expansion of the universe. Manipulating this expression, we obtain
\begin{equation}\label{eq:difeq-1}
\frac{d\log[f'(R)]}{dR}\le \frac{l^2}{1+l^2R}
\end{equation}
which can be integrated twice to give the following inequality
\begin{equation}\label{eq:fR-0}
f(R)\le A+B\left(R+\frac{l^2R^2}{2}\right)
\end{equation}
where $B$ is a positive constant, which can be set to unity
without loss of generality. Since $f'$ and $f''$ are positive, the
lagrangian is also bounded from below, i.e., $f(R)\ge A$. In
addition, according to the cosmological data, $A \equiv -2\Lambda$
must be of order a cosmological constant $2\Lambda \sim 10^{-53} $
m$^2$. We thus conclude that the gravity lagrangian at
intermediate and low scalar curvatures is bounded by
\begin{equation}\label{eq:fR-1}
-2\Lambda \le f(R)\le R-2\Lambda +\frac{l^2R^2}{2}
\end{equation}
This result shows that a lagrangian with nonlinear terms that grow
with the cosmic expansion is not compatible with the current solar
system gravitational tests, such as we argued above. Therefore,
those theories cannot represent a valid mechanism to justify the
observed cosmic speed-up.\\

\section{Summary and conclusions}

In this work we have computed the post-Newtonian limit of $f(R)$
gravities in the metric approach using a scalar-tensor
representation. This representation allows to encode the
higher-order derivatives of the metric in a self-interacting
scalar field defined by $\phi\equiv df/dR $. In this manner, the
equations of motion turn into a system of second-order equations
for the metric plus a second-order equation for the scalar field.
The post-Newtonian metric is thus characterized by several
quantities related to the scalar field, say, the intensity $V_0$
of the potential, the length scale $\sim m_\varphi ^{-1}$ of its
interaction, and the boundary values $\phi _0$ and $\dot{\phi
}_0$. Since those magnitudes are given in terms of $f$ and its
derivatives, we found the constraints given in eq.(\ref{eq:c2})
and eq.(\ref{eq:c1}) necessary to have agreement between the
predictions of these theories and the observational data in the
solar system. Those constraints show that the gravity lagrangian
$f(R)$ at relatively low curvatures is bounded from above an from
below according to eq.(\ref{eq:fR-1}). Therefore, $f(R)$ gravities
with nonlinear terms that grow with the expansion of the universe
are incompatible with observations and cannot represent a valid
mechanism to justify  the cosmic accelerated expansion rate. In
the viable models the non-linearities represent a short-range
scalar interaction, whose effect in the late-time cosmic dynamics
reduces to that of a cosmological constant and, therefore, do not
substantially modify the description provided by General Relativity
with a cosmological constant. \\

As a final remark, we want to point out that the Starobinsky model
$f(R)=R+aR^2$ \cite{STAR} besides leading to early-time inflation
and satisfying the solar system observational constraints, also
seems
compatible with CMBR observations \cite{JH-HN}.\\

\section*{Acknowledgements}
The author thanks Prof. Leonard Parker for helpful comments and
insights. Special thanks go to Prof. Jos\'{e} Navarro-Salas  for his
wise and continuous advice. I also thank H. Sanchis for his
patience and useful discussions, and L. L\'{o}pez and S. Lizardi for
hospitality during my stay in Milwaukee. This work has been
supported by a fellowship from the Regional Government of Valencia
(Spain) and the research grant BFM2002-04031-C02-01 from the
Ministerio de Educaci\'{o}n y Ciencia (Spain), and could not have been
carried out without the comprehension of Sonia G.B.

\appendix
\section{Detailed calculations}

We will take $g_{\mu \nu }\approx \eta_{\mu \nu }+h_{\mu \nu }$,
$g^{\mu \nu }\approx \eta^{\mu \nu }-h^{\mu \nu }$ and $\phi=\phi
_0+\varphi(t,x)$.  For convenience, we will rewrite the equations
of motion corresponding to the action of eq.(\ref{eq:ST}) in the
following form
\begin{eqnarray}\label{eq:Rab}
R_{\mu \nu}&=& \frac{\kappa ^2}{\phi }\left[T_{\mu \nu
}-\frac{1}{2}g_{\mu \nu }T\right]+\frac{\omega}{\phi ^2}
\partial_\mu \phi
\partial_\nu \phi +\frac{1}{\phi }\nabla_\mu \nabla_\nu \phi
+\nonumber \\ &+&\frac{1}{2\phi }g_{\mu \nu }\left[\Box \phi
+V(\phi )\right]
\end{eqnarray}
We keep the term with $\omega $ because at the same price we can
compute the post-Newtonian limit of any Brans-Dicke-like
 theory. At the end of the calculations we can
particularize to the case $\omega =0$ to obtain the desired
result. It is also useful to keep the term with $\omega $ to check
that, when the potential terms are neglected, we recover the
expected limit of Brans-Dicke
theories.\\

The expansion of the Ricci tensor around the Minkowski metric can
be written as follows
\begin{eqnarray}
R_{ij}&=& -\frac{1}{2}\nabla^2
h^{(2)}_{ij}+\frac{1}{2}\partial_i\left[h^\mu _{j,\mu
}-\frac{1}{2}h^\mu _{\mu ,j}\right]+\nonumber
\\ &+& \frac{1}{2}\partial_j\left[h^\mu _{i,\mu
}-\frac{1}{2}h^\mu _{\mu ,i}\right]
\end{eqnarray}
\begin{eqnarray}
 R_{0j}&=& -\frac{1}{2}\nabla^2
h^{(3)}_{0j}+\frac{1}{2}\partial_j\left[h^\mu _{0,\mu
}-\frac{1}{2}h^\mu _{\mu ,0}\right]+\nonumber
\\ &+&\frac{1}{2}\partial_0\left[h^\mu _{j,\mu
}-\frac{1}{2}h^\mu _{\mu ,j}\right]
\end{eqnarray}
\begin{eqnarray}
 R_{00}&=&
-\frac{1}{2}\nabla^2\left[h^{(4)}_{00}+\frac{(h^{(2)}_{00})^2}{2}\right]+\nonumber
\\ &+&\partial_0\left[h^\mu
_{0,\mu }-\frac{1}{2}h^\mu _{\mu ,0}+\frac{1}{2}h^{(2)}_{00,0}\right]+\nonumber\\
&+&\frac{1}{2}\left[h^\mu _{j,\mu }-\frac{1}{2}h^\mu _{\mu
,j}\right]\partial^jh^{(2)}_{00}+\nonumber
\\ &+&\frac{1}{2}h^{(2)}_{00}\nabla^2h^{(2)}_{00}+\frac{1}{2}h^{(2)ij}\partial_i\partial_j
h^{(2)}_{00}
\end{eqnarray}
where all the indices are raised and lowered with the Minkowski
metric. Assuming a perfect fluid, the elements on the right hand
side of eq.(\ref{eq:Rab}) are given, up to the necessary order, by
\begin{equation}
\tau _{ij}= \frac{\kappa^2\rho }{2\phi _0}\delta _{ij}+\rho O(v^2)
\end{equation}

\begin{equation}
\tau _{0j}=-\frac{\kappa ^2}{\phi _0}\rho v_j+\rho O(v^3)
\end{equation}

\begin{equation}
\tau_{00}=\frac{\kappa ^2\rho }{2\phi _0}\left[1+\Pi
+2v^2-\left(h^{(2)}_{00}+\frac{\varphi^{(2)} }{\phi
_0}\right)+\frac{3P}{\rho }\right]+\rho O(v^4)
\end{equation}
where
\begin{equation}
\tau_{\mu \nu }\equiv \frac{\kappa ^2}{\phi }\left[T_{\mu \nu
}-\frac{1}{2}g_{\mu \nu }T\right]
\end{equation}
We can also define the contribution due to the scalar field as
\begin{equation}\label{eq:tauphi}
\tau^{\phi}_{\mu \nu }\equiv \frac{\omega}{\phi ^2} \partial_\mu
\phi
\partial_\nu \phi +\frac{1}{\phi }\nabla_\mu \nabla_\nu \phi
+\frac{1}{2\phi }g_{\mu \nu }\left[\Box \phi +V(\phi )\right]
\end{equation}
Its components are
\begin{equation}
\tau^{\phi}_{ij}= \partial_i\partial_j\left(\frac{\varphi^{(2)}
   }{\phi _0}\right)+\frac{\delta _{ij}}{2\phi
_0}\left[V_0-\ddot{\phi }_0+\nabla^2\varphi^{(2)}\right]
\end{equation}

\begin{eqnarray}
\tau^{\phi}_{0j}&=& \frac{1}{2}\partial_j\left[2\omega
\frac{\dot{\phi}_0}{\phi _0}\frac{\varphi^{(2)} }{\phi
_0}+\frac{\dot{\phi}_0}{\phi
_0}h^{(2)}_{00}+\frac{\dot{\varphi}^{(2)}
}{\phi_0}\right]+\nonumber\\
&+&\frac{1}{2}\partial_0\partial_j\left(\frac{\varphi^{(2)} }{\phi
_0}\right)+\frac{1}{2}\frac{\dot{\phi}_0}{\phi
_0}\partial_j\left(\frac{\varphi^{(2)} }{\phi _0}\right)
\end{eqnarray}

\begin{eqnarray}
\tau^{\phi}_{00}&=& \partial_0\left[2\omega
\frac{\dot{\phi}_0}{\phi _0}\frac{\varphi^{(2)} }{\phi
_0}+\frac{\dot{\phi}_0}{\phi
_0}h^{(2)}_{00}+\frac{\dot{\varphi}^{(2)}
}{\phi_0}\right]+\nonumber\\&+&\frac{1}{2}\frac{\ddot{\varphi}^{(2)}}{\phi
_0}+\frac{1}{2}\partial^k
h^{(2)}_{00}\partial_k\left(\frac{\varphi^{(2)}
}{\phi _0}\right)+\nonumber \\
&+&\omega \left(\frac{\dot{\phi}_0 }{\phi
_0}\right)^2\left[1+2\left(\frac{\varphi^{(2)} }{\phi
_0}\right)\right]+\nonumber\\&+&\left(\frac{\ddot{\phi}_0 }{\phi
_0}\right)\left[\frac{3}{2}-\left(\frac{3}{2}+2\omega\right)\left(\frac{\varphi^{(2)}
}{\phi _0}\right)-h^{(2)}_{00}\right]+\nonumber\\
&+& \left(\frac{\dot{\phi}_0 }{\phi
_0}\right)\left[\left(\frac{\dot{\phi}_0 }{\phi
_0}\right)h^{(2)}_{00}
   +\left(\frac{\dot{\varphi}^{(2)} }{\phi
   _0}\right)-h^{(2)}_{00,0}-\right.\nonumber\\ &-&\left.\frac{1}{2}\left(h^\mu _{0,\mu
   }-\frac{1}{2}h^\mu _{\mu ,0}\right)\right]-\nonumber \\
&-& \frac{1}{2\phi
_0}\left[V_0\left(1-h^{(2)}_{00}-\frac{\varphi^{(2)} }{\phi
_0}\right)+\varphi^{(2)}
V_0'\right]+\nonumber\\&+&\frac{1}{2}\left[h^\mu _{k,\mu
   }-\frac{1}{2}h^\mu _{\mu ,k}\right]\partial^k\left(\frac{\varphi^{(2)}
   }{\phi _0}\right)+\nonumber \\
&+&
\frac{1}{2}\left[h^{(2)}_{00}+h^{(2)}_{[ij]}+\frac{\varphi^{(2)}
}{\phi
   _0}-1\right]\nabla^2\left(\frac{\varphi^{(2)}
   }{\phi _0}\right)
\end{eqnarray}
Using the gauge conditions
\begin{equation}
h^\mu _{k,\mu }-\frac{1}{2}h^\mu _{\mu
,k}=\frac{\partial_k\varphi^{(2)} }{\phi _0}
\end{equation}

\begin{eqnarray}
 h^\mu _{0,\mu
}-\frac{1}{2}h^\mu _{\mu ,0}&=&\left(2\omega\frac{\dot{\phi}_0
}{\phi _0}\frac{\varphi^{(2)} }{\phi _0}+\frac{\dot{\phi}_0 }{\phi
_0}h^{(2)}_{00}+\frac{\dot{\varphi}^{(2)} }{\phi
_0}\right)-\nonumber
\\ &-&\frac{1}{2}h^{(2)}_{00,0}
\end{eqnarray}
the equations of motion boil down to
\begin{equation}
-\frac{1}{2}\nabla^2[h^{(2)}_{ij}+\delta _{ij}\frac{\varphi^{(2)}
}{\phi _0}]=\frac{\delta _{ij}}{2\phi _0}\left[\kappa ^2\rho
+V_0-\ddot{\phi }_0\right]
\end{equation}

\begin{equation}
-\frac{1}{2}\nabla^2h^{(3)}_{0j}-\frac{1}{4}h^{(2)}_{00,0j}=-\frac{\kappa
^2 }{\phi_0 }\rho v_j
\end{equation}

\begin{eqnarray}
-\frac{1}{2}\nabla^2\left[h^{(4)}_{00}-\frac{\varphi^{(4)} }{\phi
_0}+\frac{(h^{(2)}_{00})^2}{2}+\frac{1}{2}\left(\frac{\varphi^{(2)}
}{\phi _0}\right)^2\right]&=& \nonumber\\ \frac{\kappa ^2\rho
}{2\phi _0}\left[1+\Pi +2v^2+h^{(2)}_{[ij]}-\frac{\varphi^{(2)}
}{\phi
_0}+\frac{3P}{\rho }\right] &+&\nonumber\\
+\frac{\ddot{\varphi}^{(2)}}{\phi _0}+
\omega\left(\frac{\dot{\phi}_0}{\phi_0}\right)^2\left[1+h^{(2)}_{00}+h^{(2)}_{[ij]}\right]&-&\nonumber\\
\frac{1}{2\phi_0}\left[V_0\left(1+h^{(2)}_{[ij]}-\frac{\varphi^{(2)}
}{\phi _0}\right)+\varphi^{(2)} V_0'\right]&+&\nonumber
\\\frac{\ddot{\phi
}_0}{\phi
_0}\left[\frac{3}{2}\left(1+h^{(2)}_{[ij]}-\frac{\varphi^{(2)}
}{\phi _0}\right)+\frac{h^{(2)}_{00}}{2}-2\omega
\frac{\varphi^{(2)} }{\phi _0}\right]&+&\nonumber\\
\frac{\dot{\phi }_0}{2\phi _0}\left[2\omega\frac{\dot{\phi}_0
}{\phi _0}\frac{\varphi^{(2)} }{\phi _0}+\frac{\dot{\phi}_0 }{\phi
_0}h^{(2)}_{00}+\frac{\dot{\varphi}^{(2)} }{\phi_0}-
\frac{1}{2}h^{(2)}_{00,0}\right]
\end{eqnarray}
where $h_{[ij]}$ simply states the relation $h_{ij}=\delta
_{ij}h_{[ij]}$. The equation for the scalar field is given by
\begin{eqnarray}
\left(\nabla^2-m_\varphi ^2\right)\left[\varphi^{(4)}
-\frac{1}{2}\frac{(\varphi^{(2)})^2}{\phi _0}\right]=-\frac{\kappa
^2\rho }{3+2\omega }\left[1+\Pi\right. &-& \nonumber\\
\left.\frac{3P}{\rho }+h^{(2)}_{[ij]}-\frac{\varphi^{(2)} }{\phi
_0}\right]+\ddot{\varphi}^{(2)}+m_\varphi ^2\varphi^{(2)}
h^{(2)}_{[ij]} &-&\nonumber
\\-\dot{\phi }_0\left[2\omega\frac{\dot{\phi}_0 }{\phi
_0}\frac{\varphi^{(2)} }{\phi _0}+\frac{\dot{\phi}_0 }{\phi
_0}h^{(2)}_{00}+\frac{\dot{\varphi}^{(2)} }{\phi
_0}-\frac{1}{2}h^{(2)}_{00,0}\right]&+&\nonumber
\\ \frac{(\varphi^{(2)})^2}{2\phi _0}\left[\frac{\phi
_0^2V_0'''}{3+2\omega }-\frac{m_\varphi ^2}{2}\right]
\end{eqnarray}
where we have defined
\begin{equation}
m_\varphi ^2=\frac{\phi _0V_0''-V_0'}{3+2\omega}
\end{equation}
The solutions are formally given by
\begin{equation}
\frac{\varphi^{(2)} (t,x)}{\phi _0}= \frac{\kappa ^2}{4\pi \phi
_0}\int d^3x' \frac{\rho
(t,x')}{|x-x'|}\frac{F(|x-x'|)}{(3+2\omega) }
\end{equation}

\begin{eqnarray}
 h^{(2)}_{00}(t,x)&=&
\frac{\kappa ^2}{4\pi \phi _0}\int d^3x'
\frac{\rho(t,x')}{|x-x'|}\left[1+\frac{F(|x-x'|)}{(3+2\omega)
}\right]+\nonumber \\ &+&\left[V_0-3 \frac{\ddot{\phi }_0}{\phi
_0}-2\omega \left(\frac{\dot{\phi }_0}{\phi_0}\right)^2
\right]\frac{|x-x_c|^2}{6}
\end{eqnarray}

\begin{eqnarray}
h^{(2)}_{ij}(t,x)&=& \left(\frac{\kappa ^2}{4\pi\phi _0 }\int
d^3x' \frac{\rho
(t,x')}{|x-x'|}\left[1-\frac{F(|x-x'|)}{(3+2\omega)
}\right]-\right.\nonumber \\ &-&\left.\frac{(V_0-\ddot{\phi
}_0)}{\phi _0}\frac{|x-x_c|^2}{6}\right)\delta _{ij}
\end{eqnarray}

\begin{eqnarray}
h_{0j}^{(3)}(t,x)&=&-\frac{\kappa ^2}{4\pi \phi_0}\int
d^3x'\frac{2\rho (t,x')v'_j}{|x-x'|}+\nonumber \\ &+&
\frac{1}{4\pi }\int d^3x'\frac{h_{00,0j}^{(2)}}{2|x-x'|}
\end{eqnarray}

\begin{eqnarray}
h_{00}^{(4)}(t,x)&=& \frac{1}{4\pi }\int d^3x'
\frac{1}{|x-x'|}\left[\frac{\ddot{\varphi }^{(2)}}{\phi
_0}+\right.\nonumber \\ &+& \left.\frac{\dot{\phi }_0}{\phi
_0}\left(2\omega\frac{\dot{\phi}_0 }{\phi _0}\frac{\varphi^{(2)}
}{\phi _0}+\frac{\dot{\phi}_0 }{\phi
_0}h^{(2)}_{00}+\frac{\dot{\varphi}^{(2)} }{\phi
_0}-\frac{1}{2}h^{(2)}_{00,0}\right)\right]+\nonumber\\
&+&\frac{\kappa^2}{4\pi \phi _0}\int d^3x' \frac{\rho
(t,x')}{|x-x'|}\left[\left(1+\Pi +h_{[ij]}^{(2)}-\frac{\varphi
^{(2)}}{\phi _0}\right)\right.\times\nonumber
\\&& \left.\times\left(1+\frac{F(|x-x'|)}{3+2\omega}\right)-\right.\nonumber \\ &-&
\left.\frac{3P}{\rho }\left(1-\frac{F(|x-x'|)}{3+2\omega }\right)+2v^2\right]+\nonumber \\
&+& \frac{\omega }{\pi }\left(\frac{\dot{\phi}_0}{\phi
_0}\right)^2\int
d^3x'\frac{[h_{00}^{(2)}+h_{[ij]}^{(2)}]}{|x-x'|}+\frac{(h_{00}^{(2)})^2}{2}-\nonumber\\
&-&\frac{1}{4\pi \phi _0}\int
d^3x'\frac{\left[V_0\left(h_{[ij]}^{(2)}-\frac{\varphi
^{(2)}}{\phi _0}\right)+\varphi ^{(2)}V_0'\right]}{|x-x'|}+\nonumber\\
&+&\frac{1}{4\pi }\int d^3x'\frac{\left(\frac{\ddot{\phi }_0}{\phi
_0}\right)}{|x-x'|}\left[3\left(h^{(2)}_{[ij]}-\frac{\varphi^{(2)}
}{\phi _0}\right)+\right. \nonumber \\ &+&
\left.h^{(2)}_{00}-4\omega \frac{\varphi^{(2)} }{\phi
_0}\right]-\nonumber \\
&-&\frac{1}{4\pi }\int
d^3x'\frac{F(|x-x'|)}{|x-x'|}\left[m_\varphi ^2\varphi^{(2)}
h^{(2)}_{[ij]}+\right.\nonumber\\
&+&\left.\frac{(\varphi^{(2)})^2}{2\phi _0}\left(\frac{\phi
_0^2V_0'''}{3+2\omega }-\frac{m_\varphi ^2}{2}\right)\right]
\end{eqnarray}

\begin{eqnarray}
\frac{\varphi^{(4)}(t,x)}{\phi _0} &=&
\frac{1}{2}\left(\frac{\varphi^{(2)}}{\phi
_0}\right)^2+\frac{\kappa ^2}{4\pi \phi _0}\int d^3x'\frac{\rho
(t,x')}{|x-x'|}\frac{F(|x-x'|)}{(3+2\omega
)}\times\nonumber\\&&\times\left[1+\Pi -\frac{3P}{\rho
}+h^{(2)}_{[ij]}-\frac{\varphi^{(2)}
}{\phi _0}\right]-\nonumber\\
&-&\frac{1}{4\pi }\int d^3x'\frac{F(|x-x'|)}{|x-x'|}\left[
\frac{\ddot{\varphi}^{(2)}}{\phi _0}
-\right.\nonumber\\&-&\left.\frac{\dot{\phi }_0}{\phi
_0}\left(2\omega\frac{\dot{\phi}_0 }{\phi _0}\frac{\varphi^{(2)}
}{\phi _0}+\frac{\dot{\phi}_0 }{\phi
_0}h^{(2)}_{00}+\frac{\dot{\varphi}^{(2)} }{\phi
_0}-\frac{1}{2}h^{(2)}_{00,0}\right)\right]\nonumber\\
&-& \frac{1}{4\pi }\int d^3x'\frac{F(|x-x'|)}{|x-x'|}\left[
m_\varphi ^2\frac{\varphi^{(2)}}{\phi
_0}h^{(2)}_{[ij]}+\right.\nonumber\\&+&\left.\frac{1}{2}\left(\frac{\varphi^{(2)}}{\phi_0}\right)^2\left(\frac{\phi
_0^2V_0'''}{3+2\omega }-\frac{m_\varphi ^2}{2}\right)\right]
\end{eqnarray}
where the function $F(|x-x'|)$ denotes
\begin{eqnarray}
F(|x-x'|)&=&\left\{\begin{array}{cc}
            e^{-m_\varphi |x-x'|} & \makebox{ if } \ m_\varphi
            ^2>0\\ \vspace{0.25cm} \\
            \cos (m_\varphi|x-x'|)  & \makebox{ if } \ m_\varphi ^2<0
            \end{array}\right.
\end{eqnarray}

\end{document}